\documentclass[conference]{IEEEtran}
\IEEEoverridecommandlockouts

\usepackage{cite}
\usepackage{amsmath,amssymb,amsfonts}
\usepackage{graphicx}
\usepackage{textcomp}
\usepackage{xcolor}
\usepackage{booktabs}
\usepackage{multirow}
\usepackage{algorithm}
\usepackage{algpseudocode}
\usepackage{dblfloatfix}   
\algrenewcommand\algorithmicrequire{\textbf{Input:}}
\algrenewcommand\algorithmicensure{\textbf{Output:}}
\usepackage{enumitem}
\usepackage{microtype}
\usepackage{colortbl}
\usepackage{float}
\usepackage{needspace}
\usepackage{etoolbox}
\usepackage{placeins}
\definecolor{groupBbg}{RGB}{253,235,235}

\newcommand{\Venv}{V_{\mathrm{env}}}
\newcommand{\Vsig}{V_{\mathrm{sig}}}
\newcommand{\Vcall}{V_{\mathrm{call}}}
\newcommand{\Vstruct}{V_{\mathrm{struct}}}

\newcommand{\REnv}{\textsc{Env\_Mismatch}}
\newcommand{\RCall}{\textsc{Call\_Mismatch}}
\newcommand{\RSig}{\textsc{Sig\_Mismatch}}
\newcommand{\RStruct}{\textsc{Struct\_Mismatch}}

\newboolean{showcomments}
\setboolean{showcomments}{true} 
\ifthenelse{\boolean{showcomments}}
{\newcommand{\nb}[2]{
		\fcolorbox{gray}{yellow}{\bfseries\sffamily\scriptsize#1}
		{$\blacktriangleright$#2$\blacktriangleleft$}
	}
	
}
{\newcommand{\nb}[2]{}
	 
}

\begin{document}
\allowdisplaybreaks

\title{An Approach for Embedding-Guided Function Reuse Detection in Embedded C Software}


\author{
\IEEEauthorblockN{A A Talha Talukder\IEEEauthorrefmark{1},
Omar Alam\IEEEauthorrefmark{1},
Akramul Azim\IEEEauthorrefmark{2}}
\IEEEauthorblockA{\IEEEauthorrefmark{1}Trent University, Peterborough, Ontario, Canada\\
Email: \{talha, omaralam\}@trentu.ca}
\IEEEauthorblockA{\IEEEauthorrefmark{2}Department of Electrical, Computer and Software Engineering, Ontario Tech University, Ontario, Canada\\
Email: akramul.azim@ontariotechu.ca}
}

\maketitle
\begin{abstract}
Reusing embedded software functions across products is
economically valuable but technically difficult: the same
functionality implemented for two different microcontroller
platforms can be entirely incompatible at the hardware level,
even when the functions score above 0.90 cosine similarity and
both pass SonarQube quality checks.
Static analysis tools were designed to measure code quality, not
hardware-domain compatibility, and have no model of peripheral
interfaces, hardware abstraction layer (HAL) dependencies, or
register-map constraints.
This paper presents a domain-aware retrieval-augmented generation
(RAG) pipeline for embedded C software reuse detection that
addresses the hardware-compatibility gap directly.
The pipeline enriches each function by extracting its existing inline comments, call-graph context, and a project README before embedding it
with eight backbone models (MiniLM, MPNet, BGE, E5,
GraphCodeBERT, OpenAI text-embedding-3-small, LLaMA 3 8B,
StarCoder2 3B) acting as feature extractors.
Four hardware-compatibility validators---covering peripheral token
overlap, parameter count parity, call-graph dependency overlap,
and structural branching pattern---filter candidates directly
in the retrieval stack.
Evaluated on six public embedded C software projects (184
functions, 4,815 above-plateau pairs), the pipeline reveals that
SonarQube produces a 93.6\% false-positive rate as a reuse
filter, with 83.5\% of failures caused by hardware-environment
mismatches that static analysis cannot detect.
Manual verification of 40 rejected pairs confirms 97.5\%
validator accuracy, and a diagnostic rule-injection variant
identifies the dominant failure categories
(McNemar chi-squared~=~294.0, p~$<$~0.001).

\end{abstract}


\section{Introduction}
\label{sec:intro}

Embedded software reuse has long been recognised as a strategy
to reduce development cost and accelerate product cycles in the
embedded systems industry~\cite{maruf2022reuse,maruf2024feamod}.
When a company produces multiple products on the same
microcontroller family, e.g., a microwave oven controller and an
electric water heater, both running on Texas Instruments Tiva~C,
it is natural to ask whether software functions developed for one
product can be safely reused in another. However, designing
software to facilitate modular reuse remains a challenging
software engineering problem~\cite{ThimmegowdaASADMMK14}.
Reusing a well-tested sensor driver or peripheral-initialisation
routine would reduce duplicated effort and inherit bug fixes
already validated in the source project.

However, embedded software reuse is qualitatively harder than
reuse in general-purpose software~\cite{maruf2022reuse,clements2020halucinator}.
Unlike a sorting algorithm or a string formatter, an embedded
function's correctness is tightly coupled to the specific
hardware it controls: its target microcontroller, its peripheral
register map, and the vendor hardware abstraction layer (HAL) it
depends on.
A function that drives GPIO~Port~F through TivaWare's
\texttt{GPIO\_writePort()} cannot be substituted for a function
that drives GPIO~Port~C through ARM's \texttt{HAL\_GPIO\_WritePin()},
even when both implement identical logical behaviour.
This coupling creates three systematic obstacles~\cite{clements2020halucinator,p2im2020,pararehosting2021}.
\textit{First, tight hardware coupling:} a software function's
correctness depends on the exact peripheral it controls, not
merely its logic.
\textit{Second, HAL fragmentation:} vendor layers---TivaWare,
ARM CMSIS, ARM~LL, ESP-IDF, Zephyr---provide overlapping but
incompatible APIs, so functions with identical intent are not
interchangeable across projects.
\textit{Third, documentation scarcity:} embedded codebases carry
terse inline comments and product-level READMEs, depriving
code-embedding models of the natural-language signals on which
they were trained~\cite{bansal2023fcg,lomshakov2024proconsul}.

The state of the art offers two kinds of tools for identifying
reuse candidates: static analysis tools such as SonarQube~\cite{lenarduzzi2020sonar,sadowski2018google}, and embedding-based code-similarity search~\cite{feng2020codebert,guo2021graphcodebert}.
SonarQube is widely used as a quality gate in industrial
development pipelines~\cite{sadowski2018google}: functions that
pass its checks are considered clean enough to consider for
reuse.
Embedding-based search scores function pairs by cosine
similarity over learned code representations, surfacing
structurally and semantically similar candidates.
Both tools are effective in their intended domains---but neither
was designed to reason about hardware-domain compatibility.
SonarQube checks cyclomatic complexity, null-pointer risks, and
coding standards~\cite{lenarduzzi2020sonar}; it has no model of
which peripherals a function touches, which HAL it depends on,
or what register space it occupies.
Embedding models assign high similarity to functions that share
logical structure and natural-language comments, regardless of
whether they target the same or different hardware
interfaces~\cite{muennighoff2023mteb}.
As a consequence, two functions that are entirely
hardware-incompatible can simultaneously pass SonarQube and
score above 0.90 cosine similarity---and no existing tool will
flag this mismatch before the ported code fails at hardware test.

Despite this known limitation, no prior work has empirically
measured how severely SonarQube's quality-only model fails as a
reuse-compatibility filter for embedded C software.
We address this gap with a domain-aware retrieval-augmented
generation (RAG) pipeline for embedded C reuse detection that
integrates hardware-compatibility validation directly into the
retrieval stack, rather than checking compatibility post-hoc.
The pipeline enriches each software function with inline
comments, call-graph context, and a project README into a single
document DOC($f$), which is then embedded using eight backbone
models spanning compact encoders~\cite{wang2020minilm,song2020mpnet}, retrieval-tuned bi-encoders~\cite{chen2024bge,wang2022e5}, a code transformer~\cite{guo2021graphcodebert}, and large language models~\cite{grattafiori2024llama,lozhkov2024starcoder}.
Four hardware-compatibility validators---$\Venv$ (peripheral
token overlap), $\Vsig$ (parameter count parity), $\Vcall$
(call-graph dependency overlap), and $\Vstruct$ (structural
branching pattern)---filter candidates directly in the retrieval
stack.
An unsupervised plateau analysis calibrates a per-model cosine
threshold without requiring labelled data.
A basic RAG run is performed first; the validator failures from
that run are clustered to derive a rule library~$\mathcal{L}$,
which a dynamic-rule RAG variant then injects to expose which
failure categories drive retrieval quality.

We evaluate the pipeline on six public embedded C software
repositories across three domains (184 functions, 4,815
above-plateau pairs).
The central empirical finding is that SonarQube produces a
\textbf{93.6\% false-positive rate} when used as a reuse
filter: of 1,494 pairs it approves as quality-clean, 1,399 are
rejected by our domain validators, with 83.5\% of failures
caused by hardware-environment mismatches that static analysis
cannot detect.
Manual verification of 40 randomly selected rejected pairs
confirms 97.5\% validator accuracy.

The main contributions of this paper are as follows:
\begin{itemize}[leftmargin=*,topsep=2pt,itemsep=2pt]
  \item We propose a domain-aware RAG pipeline for embedded C
    software reuse detection that integrates four
    hardware-compatibility validators directly into the retrieval
    stack, enabling hardware-aware candidate filtering without
    labelled training data (Section~\ref{sec:method}).
  \item We introduce four embedded-domain validators---$\Venv$,
    $\Vsig$, $\Vcall$, and $\Vstruct$---that together capture
    peripheral identity, interface shape, dependency burden, and
    structural branching pattern, providing explainable rejection
    signals for each incompatible pair (Section~\ref{sec:validators}).
  \item We provide the first empirical quantification of
    SonarQube's false-positive rate as a reuse-compatibility
    filter for embedded C software: 93.6\%, with 83.5\% of
    failures attributable to hardware-environment mismatches
    that static analysis is structurally unable to detect
    (Section~\ref{sec:motivation}).
  \item We evaluate the pipeline across eight embedding
    backbones spanning four architectural families, demonstrating
    that the 93.6\% false-positive finding holds across all
    models (per-model rates 89.8\%--97.3\%), and identify
    BGE-small, LLaMA~3 8B, and MPNet as the top-performing
    backbones for this task (Section~\ref{sec:rq2}).
  \item We develop a dynamic-rule RAG diagnostic variant that
    exposes hardware-token and call-graph mismatch as the
    dominant retrieval failure categories, confirmed by a
    statistically significant McNemar test
    ($\chi^2=294.0$, $p<0.001$, Section~\ref{sec:rq4}).
\end{itemize}

The remainder of this paper is organised as follows.
Section~\ref{sec:related} surveys related work.
Section~\ref{sec:motivation} presents the empirical motivation.
Section~\ref{sec:method} describes the methodology and algorithms.
Section~\ref{sec:experiments} reports the experimental setup.
Section~\ref{sec:results} answers each research question.
Section~\ref{sec:discussion} discusses implications and threats.
Section~\ref{sec:conclusion} concludes.

\section{Related Work}
\label{sec:related}
In contrast to existing approaches that primarily rely on code similarity or static analysis, we propose a domain-aware RAG pipeline for embedded C software reuse detection that integrates hardware-compatibility validators for explainable candidate filtering. Our approach captures hardware-driven incompatibilities, enabling more reliable reuse detection without requiring labelled training data. Below, we discuss some of the related work to our approach.

\noindent\textbf{Embedded software reuse.}
Maruf et al.~\cite{maruf2022reuse} extract reusable functions
via static call-graph analysis; Talukder et
al.~\cite{talukder2025llm} use LLMs for feature extraction;
FeaMod~\cite{maruf2024feamod} targets modularity.
None quantify static-tool false-positive rates or propose
hardware-compatibility validators.
AutoFirm~\cite{autofirm2024} finds that 67.3\% of IoT vendors
fail to update reused libraries, confirming that compatibility
checking is largely absent from current practice.

\noindent\textbf{Static analysis limitations.}
Lenarduzzi et al.~\cite{lenarduzzi2020sonar,lenarduzzi2020rules}
show SonarQube rules have statistically significant but small
effects on fault-proneness across 33 Apache projects.
Sadowski et al.~\cite{sadowski2018google} establish 10\% as the
industrial adoption floor above which FP rates collapse developer
trust.
Charoenwet et al.~\cite{charoenwet2024sast} report $\geq$76\%
of SAST warnings in C/C++ vulnerability detection are
irrelevant---the closest published analog to our 93.6\% finding.
Cui et al.~\cite{cui2024fnfp} catalogue SonarQube FP root
causes but do not include hardware-environment mismatch.
Johnson et al.~\cite{johnson2013} identify false positives as
the dominant barrier to ASAT adoption.

\noindent\textbf{Code quality vs.\ reusability.}
Papamichail et al.~\cite{papamichail2019reuse} show that standard
static metrics (complexity, coupling, cohesion) do not predict
reuse rates; Mehboob et al.~\cite{mehboob2021reuse} confirm
they measure quality proxies rather than portability.
In embedded software the disconnect is sharper: a well-structured
HAL function couples tightly to a specific peripheral and is
therefore \emph{less} portable than a structurally messier
hardware-free helper.

\noindent\textbf{Code clone and similarity detection.}
SourcererCC~\cite{sajnani2016sourcecc} demonstrates that
token-bag Jaccard scales to 250~million lines for Types~1--3
clones, underpinning our $\Venv$ and $\Vstruct$ validators.
The contribution is not the Jaccard mechanic but the
vocabulary: hardware-identifying tokens ($\Pi_\mathrm{hw}$)
turn a generic similarity measure into a domain-aware
compatibility check.
CCGraph~\cite{zou2020ccgraph} and FA-AST~\cite{wang2020faast}
apply graph-neural networks to structural similarity,
motivating our lightweight branching-pattern fingerprint.

\noindent\textbf{Pre-trained models for code.}
CodeBERT~\cite{feng2020codebert} and
GraphCodeBERT~\cite{guo2021graphcodebert} advance code search
on mainstream languages; BGE~\cite{chen2024bge} and
E5~\cite{wang2022e5} optimise retrieval via contrastive
pretraining.
Muennighoff et al.~\cite{muennighoff2023mteb} show similarity
distributions vary by up to 0.3 cosine units across backbones,
motivating our per-model plateau calibration rather than a fixed
global threshold.

\noindent\textbf{RAG for code.}
CoCoMIC~\cite{ding2024cocomic} reports $+33.94\%$ exact match
from cross-file call-graph context, directly motivating our
DOC($f$) construction.
DocPrompting~\cite{zhou2023docprompting} and
ProConSuL~\cite{lomshakov2024proconsul} show that retrieved
documentation improves code generation and summarisation,
motivating our README enrichment.
FirmUp~\cite{firmup2018} establishes that call-graph context
is necessary for accurate firmware function matching.
Asteria-Pro~\cite{asteriapro2023} achieves 91.65\% precision
on vulnerable IoT function detection by combining deep-learning
similarity with explicit domain knowledge---the closest
architectural analog to our four-validator stack.

CheckList~\cite{ribeiro2020checklist} and HANS~\cite{mccoy2019hans} establish behavioral rule-injection
as a first-class research contribution for exposing systematic
model failures.
Errudite~\cite{wu2019errudite} formalises error analysis as a
primary output.
Our dynamic-rule RAG variant follows this tradition: its value
lies in diagnosing failure categories, not in improving headline
metrics.

\section{Motivation: SonarQube as a Reuse Filter}
\label{sec:motivation}

SonarQube checks cyclomatic complexity, null-pointer risks, and
coding standards---none of which capture whether two functions
touch compatible hardware peripherals.
For example, when a function in one project uses \texttt{GPIO\_writePort()} and
a function in another project uses \texttt{HAL\_GPIO\_WritePin()},
both can pass SonarQube cleanly even though they target entirely
different register maps and HAL layers.
To make this gap concrete, we evaluated all 4,815 above-plateau
function pairs from six embedded C repositories using both
SonarQube and our domain validators, classifying each pair into
one of four groups (Table~\ref{tab:groups}).

\begin{table}[t]
\caption{Classification of 4,815 above-plateau pairs
  by SonarQube and validator outcome.}
\label{tab:groups}
\begin{center}
\footnotesize
\begin{tabular}{c c c r r}
\hline
\textbf{Group} & \textbf{Sonar} & \textbf{Validators} &
  \textbf{Count} & \textbf{\%} \\
\hline
A & Pass & Pass &    95 &  1.97 \\
\rowcolor{groupBbg}
B & Pass & Fail & 1,399 & 29.06 \\
C & Fail & Pass &    33 &  0.69 \\
D & Fail & Fail & 3,288 & 68.29 \\
\hline
\multicolumn{3}{l}{Total} & 4,815 & 100.0 \\
\hline
\end{tabular}
\end{center}
\end{table}

Table~\ref{tab:groups} classifies all 4,815 pairs into four groups based on two independent judgments: SonarQube's quality verdict (pass/fail) and our validators' compatibility verdict (pass/fail). Group~A pairs are approved by both — these are the genuine reuse candidates. Group~D pairs are rejected by both — SonarQube and our validators agree they are unsuitable. Groups~B and~C are the disagreements. Group~C (33 pairs, 0.69\%) represents cases where SonarQube flags code quality issues but our validators confirm hardware compatibility — a minor discrepancy. Group~B (1,399 pairs, 29.06\%) is the critical case: SonarQube approves these pairs as quality-clean, yet our hardware-compatibility validators reject every one of them. These are the false positives — pairs that would be mistakenly considered for reuse based on SonarQube alone.\noindent\textbf{Group~B is the false-positive problem.}
SonarQube approved $|A|+|B|=1{,}494$ pairs as quality-clean.
Our validators reject 1,399 of those.
The Sonar false-positive rate is:
\begin{equation}
  \text{Sonar FP rate} \;=\;
  \frac{|B|}{|A|+|B|}
  \;=\; \frac{1{,}399}{1{,}494}
  \;=\; 93.6\%
  \label{eq:fprate}
\end{equation}
This rate is $9.4\times$ the 10\% industrial adoption floor
identified by Sadowski et al.~\cite{sadowski2018google} and
17~percentage points above the $\geq$76\% irrelevance rate for
C/C++ vulnerability detection~\cite{charoenwet2024sast}.

Table~\ref{tab:reasons} breaks down why validators rejected
Group~B pairs.
$\Venv$ (hardware-token mismatch) is involved in
$527{+}243{+}228{+}170=1{,}168$ pairs, or \textbf{83.5\%} of
Group~B failures.
Among these, \textbf{78.4\%} of Group~B pairs have
$\Venv\text{-score}=0$---zero hardware-token overlap---yet
SonarQube passed all of them.
This confirms that static quality analysis cannot substitute for
domain compatibility checking in embedded
software~\cite{p2im2020,pararehosting2021,firmup2018}.

Table~\ref{tab:sonar_valid} reveals a counterintuitive result: of the 128 validator-approved pairs, none are entirely free of SonarQube warnings. This suggests that code quality and hardware-compatibility are not merely uncorrelated in this domain — they point in opposite directions. A well-structured, warning-free HAL driver is typically one that has been carefully optimised for a specific peripheral interface, making it more hardware-specific and therefore less portable to another project. Conversely, functions with some code-quality issues (e.g., unused variables or non-standard constructs) may be more generic in their hardware dependencies, making them more portable~\cite{papamichail2019reuse,mehboob2021reuse}.

\begin{table}[t]
\caption{Validator failure reasons for Group~B (1,399 pairs).
  $\Venv$ is involved in 83.5\% of all failures.}
\label{tab:reasons}
\begin{center}
\footnotesize
\begin{tabular}{l r r}
\hline
\textbf{Failure Reason} & \textbf{Count} & \textbf{\%} \\
\hline
Venv + Vcall                 & 527 & 37.7 \\
Venv + Vsig + Vcall          & 243 & 17.4 \\
Venv only                    & 228 & 16.3 \\
Venv + Vsig                  & 170 & 12.2 \\
Vcall only                   & 133 &  9.5 \\
Vsig + Vcall                 &  50 &  3.6 \\
Vsig only                    &  48 &  3.4 \\
\hline
\end{tabular}
\end{center}
\end{table}

\begin{table}[t]
\caption{SonarQube status of the 128 validator-approved pairs.}
\label{tab:sonar_valid}
\begin{center}
\footnotesize
\begin{tabular}{l r r}
\hline
\textbf{Sonar status} & \textbf{Count} & \textbf{\%} \\
\hline
Both functions Sonar-clean    &   0 &  0.0 \\
One function Sonar-flagged    & 107 & 83.6 \\
Both functions Sonar-flagged  &  21 & 16.4 \\
\hline
\end{tabular}
\end{center}
\end{table}

\noindent\textbf{Manual verification.}
To confirm that validator rejections reflect genuine hardware
incompatibilities rather than validator errors, we manually
inspected a random sample of 40 Group~B pairs with
($\Venv$) equals zero — meaning the two functions share no hardware-identifying tokens whatsoever (Table~\ref{tab:manual}).
\textbf{39 of 40 pairs (97.5\%) were confirmed genuinely
hardware-incompatible.}
The remaining pair (\texttt{init\_LCD} vs.\ \texttt{LCD\_INIT},
pair~30) uses different GPIO port assignments on different Tiva~C
boards; it is borderline---not directly portable without
pin-remapping.
No pair was found to be directly reusable without hardware
modification.
Critically, 12 of the 40 pairs had embedding similarity
$\geq$0.90 yet were confirmed incompatible, including
\texttt{state\_A} vs.\ \texttt{moistureSensor} (sim\,=\,0.962)
and \texttt{ADC\_Read} vs.\ \texttt{LED\_BUZZER} (sim\,=\,0.960).
These cases show that high semantic similarity is \emph{not
sufficient} for embedded reuse.


\begin{table}[t]
\caption{Representative manual-verification examples from Group~B
($V_{\text{env}}\text{-score}=0$, Sonar-pass, Validator-fail).}
\label{tab:manual}
\centering
\scriptsize
\setlength{\tabcolsep}{3pt}
\begin{tabular}{c l l c p{1.9cm}}
\hline
\textbf{\#} & \textbf{Source} & \textbf{Target} &
\textbf{Sim.} & \textbf{Incompatibility} \\
\hline
32 & \texttt{state\_A} & \texttt{moistureSensor} & 0.962 &
Microwave idle vs.\ HVAC I$^{2}$C sensor. \\
37 & \texttt{ADC\_Read} & \texttt{LED\_BUZZER} & 0.960 &
ADC input vs.\ GPIO output. \\
10 & \texttt{LCD\_write} & \texttt{Generic\_delay} & 0.931 &
LCD GPIO vs.\ SysTick delay. \\
30 & \texttt{init\_LCD} & \texttt{LCD\_INIT} & 0.713 &
Same LCD, different GPIO port (\emph{borderline}). \\
\hline
\multicolumn{5}{l}{\textbf{Summary:} 39 incompatible, 1 borderline, 0 reusable.} \\
\hline
\end{tabular}
\end{table}
\section{Methodology}
\label{sec:method}

\begin{figure}[!t]
  \centering
  \includegraphics[width=\columnwidth]{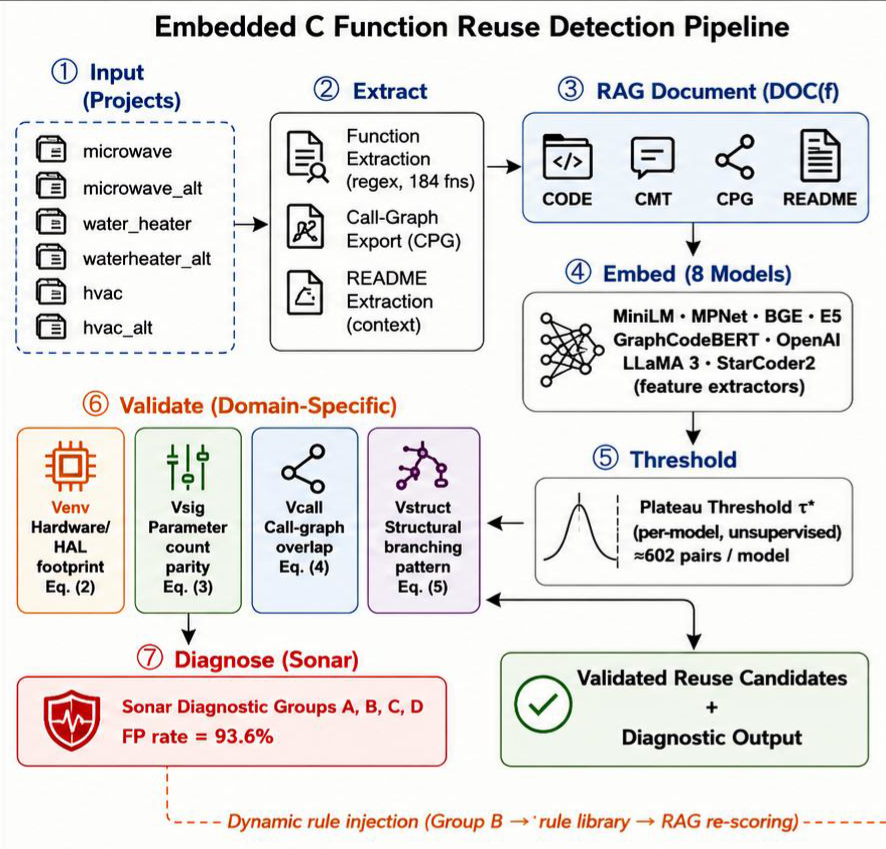}
  \caption{Seven-stage embedded C reuse detection pipeline.
    Stages: (1)~six repos, (2)~regex extraction (184 fns),
    (3)~RAG enrichment DOC($f$), (4)~eight embeddings,
    (5)~plateau threshold $\tau^*$, (6)~four validators
    ($\Venv$,$\Vsig$,$\Vcall$,$\Vstruct$), (7)~Sonar diagnostic.
    Amber dashed arc: dynamic rule injection from Group~B.}
  \label{fig:pipeline}
\end{figure}

\begin{figure}[!t]
  \centering
  \includegraphics[width=\columnwidth]{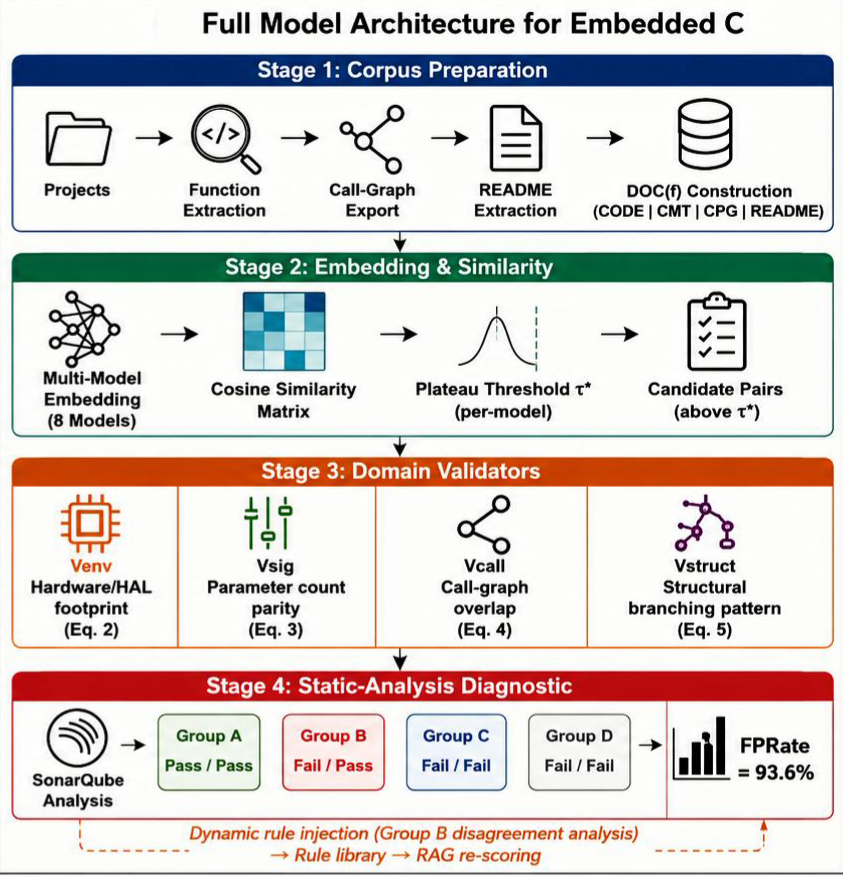}
  \caption{Full model architecture across four stages:
    corpus preparation with DOC($f$) construction,
    eight-backbone embedding and plateau calibration (Alg.~2),
    four domain validators (Eqs.~\ref{eq:venv}--\ref{eq:vstruct}),
    and SonarQube diagnostic (Groups A--D; 93.6\% FP rate).
    Amber arc: Group~B rule injection (Alg.~3).}
  \label{fig:fullmodel}
\end{figure}

\noindent\textbf{Overview.}
Fig.~\ref{fig:pipeline} shows the seven-stage pipeline and
Fig.~\ref{fig:fullmodel} shows the detailed component architecture.
The pipeline takes software repositories and an embedding model
as input and returns a validated set of reuse-compatible function
pairs together with a SonarQube diagnostic breakdown.

\subsection{Dataset and Function Extraction}
\label{sec:dataset}

We evaluated six public GitHub repositories spanning three
embedded domains, each with two independent implementations
(Table~\ref{tab:projects}).
The domain pairing allows measurement of both same-domain reuse
(e.g., microwave $\to$ microwave\_alt) and cross-domain reuse
(e.g., microwave $\to$ hvac).
A regex-based extractor isolates each top-level C function
definition (return type, name, argument list, body, line range),
yielding 184 functions in total. Each extracted function retains its original source-level
formatting and structure.

\begin{table}[t]
\caption{Firmware repositories and extracted function counts.}
\label{tab:projects}
\begin{center}
\footnotesize
\begin{tabular}{l l r}
\hline
\textbf{Project} & \textbf{Domain} & \textbf{Functions} \\
\hline
microwave        & Microwave oven controller  & 41 \\
microwave\_alt   & Microwave (alt.\ impl.)    & 35 \\
water\_heater    & Electric water heater      & 32 \\
waterheater\_alt & Water heater (alt.\ impl.) & 17 \\
hvac             & HVAC controller            & 10 \\
hvac\_alt        & HVAC (alt.\ impl.)         & 49 \\
\hline
\multicolumn{2}{l}{Total} & 184 \\
\hline
\end{tabular}
\end{center}
\end{table}

\subsection{Retrieval-Augmented Function Documents}
\label{sec:rag-doc}

Embedding only the raw function body discards two categories of
information critical for embedded reuse decisions.
First, \textit{hardware intent} is frequently expressed only in
inline comments (e.g., ``writes 0x01 to GPIO~Port~F to enable
the buzzer'')---without the comment, an embedding model sees an
assignment to an opaque register name.
Second, \textit{context} captured in the call graph determines
whether the dependency burden of porting a function is low or high.

For each software function $f$ we construct an enriched document
DOC($f$) by concatenating four text components separated by a
delimiter token:
(i)~the raw source code of $f$;
(ii)~all inline and block comments within $f$'s line range;
(iii)~a textual call-graph summary---the names of functions $f$
calls and the names of functions that call $f$; and
(iv)~a short excerpt from the project README describing the
system context (e.g., ``Microwave oven controller running on
TM4C123GH6PM with 4$\times$4 keypad, HD44780 LCD, and piezo
buzzer'').
We pass DOC($f$) to the embedding model.
The design is motivated by CoCoMIC~\cite{ding2024cocomic}
($+33.94\%$ exact match from call-graph context),
DocPrompting~\cite{zhou2023docprompting} (documentation improves
retrieval), and ProConSuL~\cite{lomshakov2024proconsul}
(project-level context aids LLM code summarisation).

\subsection{Embedding Models}
\label{sec:emb}

Table~\ref{tab:models} lists the eight backbones evaluated.
All eight act as feature extractors; they produce cosine
similarity scores over DOC($f$) pairs but do not generate or
modify code.
The multi-backbone design is essential because
Muennighoff et al.~\cite{muennighoff2023mteb} show that
similarity score distributions vary by up to 0.3 cosine units
across models, so any single-model result risks being
model-specific.
For large language models (LLaMA~3, StarCoder2) we use
mean-pooled last-layer representations; for BERT-family models
we use \texttt{[CLS]} pooling; for BGE/E5 we use the
model-default pooler.

\begin{table}[t]
\caption{Embedding backbones evaluated.}
\label{tab:models}
\begin{center}
\footnotesize
\begin{tabular}{l l p{3.0cm}}
\hline
\textbf{Key} & \textbf{Model} & \textbf{Training} \\
\hline
minilm    & MiniLM-L3        & Self-attn distillation~\cite{wang2020minilm} \\
mpnet     & MPNet            & Masked+permuted LM~\cite{song2020mpnet} \\
bge       & BGE-small-v1.5   & Contrastive retrieval~\cite{chen2024bge} \\
e5        & E5-small         & Weakly supervised~\cite{wang2022e5} \\
gcbert    & GraphCodeBERT    & Code+data-flow~\cite{guo2021graphcodebert} \\
gpt       & text-emb-3-small & OpenAI API~\cite{openai2024} \\
llama     & LLaMA 3 8B       & LLM, mean-pooled~\cite{grattafiori2024llama} \\
starcoder & StarCoder2 3B    & Code LLM~\cite{lozhkov2024starcoder} \\
\hline
\end{tabular}
\end{center}
\end{table}

\subsection{Domain-Specific Validators}
\label{sec:validators}
The four validators defined here identify Group~B pairs —
approved by SonarQube but hardware-incompatible — by checking
four orthogonal aspects of compatibility that SonarQube cannot
model, encoding the hardware-domain knowledge that pure semantic
similarity lacks.
A candidate pair $(f_s,f_t)$ is accepted only when it passes
\textit{all four} simultaneously; failure on any one is a
diagnostic signal for why that pair cannot be reused.
The validators target four orthogonal aspects of compatibility:
\begin{itemize}[topsep=2pt,itemsep=1pt,leftmargin=*]
  \item $\Venv$ --- Hardware/HAL footprint
    (Section~\ref{sec:venv})
  \item $\Vsig$ --- Parameter count parity
    (Section~\ref{sec:vsig})
  \item $\Vcall$ --- Call-footprint similarity
    (Section~\ref{sec:vcall})
  \item $\Vstruct$ --- Structural branching pattern
    (Section~\ref{sec:vstruct})
\end{itemize}
\subsubsection{$\Venv$ --- Hardware/HAL Footprint}
\label{sec:venv}
\noindent\textbf{Intuition.}
Two embedded functions are hardware-compatible only if they touch
similar peripherals --- an ADC reader and a GPIO writer share
nothing at the hardware level, regardless of how similar their
branching patterns look. We capture this via Jaccard overlap of
each function's hardware-identifying tokens.
\noindent\textbf{Definition.}
Let $H(f)$ be the set of hardware-identifying tokens in $f$'s
body, comments, and call-graph neighbourhood --- drawn from the
pattern set $\Pi_\mathrm{hw}$: \texttt{GPIO\_*}, \texttt{ADC\_*},
\texttt{UART\_*}, \texttt{SPI\_*}, \texttt{PWM\_*},
\texttt{TIMER\_*}, \texttt{HAL\_*}, \texttt{LL\_*}, vendor
register names (e.g., \texttt{PORTF}, \texttt{GPIO\_PORTC\_DATA\_R}), and HAL
function families (e.g., \texttt{HAL\_GPIO\_WritePin},
\texttt{gpio\_set\_level}). $H(f)$ answers: \textit{which
hardware peripherals does $f$ touch?}
\begin{equation}
\Venv(f_s,f_t)=
\begin{cases}
1, & H(f_s)=\emptyset \land H(f_t)=\emptyset,\\[2pt]
0, &
\begin{aligned}[t]
&\text{exactly one of } H(f_s),\\
&H(f_t)\text{ is }\emptyset,
\end{aligned}\\[2pt]
\dfrac{|H(f_s)\cap H(f_t)|}
{|H(f_s)\cup H(f_t)|},
& \text{otherwise}.
\end{cases}
\label{eq:venv}
\end{equation}
The equation covers three cases. When neither function touches
hardware, both are pure logic helpers that move freely between
projects (score~1, accepted). When exactly one does, substitution
would silently drop hardware interactions the destination may
lack (score~0, always rejected). When both do, the Jaccard score
measures the overlap of their hardware footprints, and the pair
is accepted once at least one token is shared (score~$>0$),
indicating at least partial peripheral compatibility.
\noindent\textbf{Example.}
\texttt{state\_A} has $H=\emptyset$;
\texttt{moistureSensor} has $H=\{$\texttt{ADC\_Read},
\texttt{ADC0\_BASE}, \texttt{I2C\_master\_init}$\ldots\}$.
Exactly one is empty $\Rightarrow$ Case~2 $\Rightarrow$
$\Venv=0$ $\Rightarrow$ \textbf{rejected} (sim\ =\ 0.962).
\subsubsection{$\Vsig$ --- Parameter Count Parity}
\label{sec:vsig}
\noindent\textbf{Intuition.}
Drop-in reuse requires both functions to accept the same number
of arguments --- a mismatched count breaks every call site
without rewriting callers, a heavier change than copying the
function body. This validator checks argument \textit{count}
only; type-level signature matching is left as future work.
\noindent\textbf{Definition.}
Let $\#\mathrm{args}(f)$ be the number of formal parameters.
\begin{equation}
  \Vsig(f_s,f_t) =
  1 - \frac{|\,\#\mathrm{args}(f_s) - \#\mathrm{args}(f_t)\,|}
           {\max(\#\mathrm{args}(f_s),\;\#\mathrm{args}(f_t),\;1)}
  \label{eq:vsig}
\end{equation}
The formula normalises the argument-count difference by the
maximum of the two counts, producing a score in $[0, 1]$.
Acceptance requires $\Vsig=1$, i.e., argument counts exactly
equal --- the only condition under which drop-in reuse is
possible at every call site without modifying callers. Any score
below 1 means the counts differ, and copying the function would
break call sites passing the wrong argument count. The
$\max(\ldots,1)$ guard prevents division by zero for
no-argument functions, which score $\Vsig=1$ and pass through to
the remaining validators.
\noindent\textbf{Example.}
\texttt{BUTTON\_READ(port, pin)} has 2 args;
\texttt{ReadPin(pin\_id)} has 1.
$|2-1|/\max(2,1,1)=0.5 \Rightarrow \Vsig=0.5 \Rightarrow$
\textbf{rejected}.
\subsubsection{$\Vcall$ --- Call-Footprint Similarity}
\label{sec:vcall}
\noindent\textbf{Intuition.}
A function's non-HAL dependencies form part of its portability
contract: porting $f_s$ into $f_t$'s project requires porting
every non-HAL helper $f_s$ calls that isn't already present at
the destination. The call-graph overlap measures how much of
this burden $f_s$ and $f_t$ already share; HAL calls are excluded
from $C(f)$ since they're already captured by $\Venv$, avoiding
double-counting the hardware-mismatch signal.
\noindent\textbf{Definition.}
Let $C(f)$ be the set of non-HAL functions called by $f$.
\begin{equation}
  \Vcall(f_s,f_t) =
  \begin{cases}
    1 & C(f_s){=}\emptyset \;\wedge\; C(f_t){=}\emptyset \\[2pt]
    \frac{|C(f_s)\cap C(f_t)|}{|C(f_s)\cup C(f_t)|}
      & \text{otherwise}
  \end{cases}
  \label{eq:vcall}
\end{equation}
A pair is \textbf{accepted} when $\Vcall>0$. \textit{Case~1}:
both are leaf functions (call no user-defined helpers) and carry
no transitive dependency burden, so their portability contract is
satisfied (score~$=1$), and they proceed to $\Vstruct$ for
structural screening. \textit{Case~2}: Jaccard similarity over
the non-HAL call sets --- if one function calls non-HAL helpers
and the other calls none, the score is~0 and the pair is
rejected, since porting the non-leaf function would require
transplanting all its dependencies into a destination that cannot
satisfy them.
\noindent\textbf{Examples.}
$C(\texttt{Buzzer\_ON})=C(\texttt{ArrayLED\_ON})=\{\texttt{delay\_ms}\}$:
$\Vcall=1 \Rightarrow$ \textbf{accepted}.
$C(\texttt{state\_A})=\{\texttt{setOutput}\}$,
$C(\texttt{moistureSensor})=\{\texttt{ADC\_Read},\texttt{ADC\_Configure}\}$:
$\Vcall=0 \Rightarrow$ \textbf{rejected}.
\subsubsection{$\Vstruct$ --- Structural Branching Pattern}
\label{sec:vstruct}
\noindent\textbf{Intuition.}
Two functions with very different branching structures --- a
tight register-read helper versus a multi-state event handler ---
are unlikely to be interchangeable even if they share hardware
context and call dependencies, which we capture via a count-based
comparison of branching keywords.
\noindent\textbf{Definition.}
Let $K(f)$ be the \textit{multiset} of branching keywords in
$f$'s body, drawn from
$\{$\texttt{if}, \texttt{else}, \texttt{for}, \texttt{while},
\texttt{do}, \texttt{switch}, \texttt{case}, \texttt{return},
\texttt{break}, \texttt{continue}$\}$, recording counts so that
three \texttt{if}s contribute \texttt{if}$\times$3, distinguishing
it from a single \texttt{if}.
\begin{equation}
  \Vstruct(f_s,f_t) =
  \begin{cases}
    1 & K(f_s){=}\emptyset \;\wedge\; K(f_t){=}\emptyset \\[2pt]
    \frac{|K(f_s)\cap K(f_t)|}{|K(f_s)\cup K(f_t)|}
      & \text{otherwise}
  \end{cases}
  \label{eq:vstruct}
\end{equation}
Intersection uses element-wise minimum, union element-wise
maximum. A pair is \textbf{accepted} when $\Vstruct\geq 0$
(always true) --- the multiset representation lets three
\texttt{if}s register as more committed to conditional dispatch
than one, a distinction the intersection count captures directly.
\noindent\textbf{Example.}
$K(\texttt{state\_A})=\{$\texttt{if}$\times$2,\texttt{return}$\times$1$\}$;
$K(\texttt{moistureSensor})=\{$\texttt{while}$\times$1,\texttt{return}$\times$1$\}$.
Intersection~$=1$, union~$=4$, $\Vstruct=0.25 \geq 0$
(diagnostic role, not hard rejection).

\subsection{Plateau-Based Threshold Calibration}
\label{sec:plateau}

\noindent\textbf{Why per-model calibration is needed.}
Cosine similarity is not comparable across embedding spaces.
BGE-small's contrastive pretraining produces a well-separated
distribution where 0.71 already indicates a strong match;
E5-small packs most distinct pairs near 0.85, requiring
$\sim$0.92 for equivalent retrieval
quality~\cite{muennighoff2023mteb,wang2022e5}.
A single global threshold would flood tightly-packed models with
false candidates or starve well-separated models of true matches.

\noindent\textbf{Definition.}
For each model we sweep $N=200$ cosine thresholds $\tau$
uniformly from the observed $\tau_{\min}$ to $\tau_{\max}$.
At each $\tau$ we count all pairs satisfying both the similarity
cut and all four validator conditions:

\begin{equation}
\mathrm{cnt}(\tau)=
\left|
\left\{
(f_s,f_t)\;\middle|\;
\begin{aligned}
&\mathrm{sim}(f_s,f_t)\ge\tau,\;
\Venv>0,\;
\Vsig=1,\\
&\Vcall>0,\;
\Vstruct\ge0
\end{aligned}
\right\}
\right|
\label{eq:plateau_cond}
\end{equation}
As $\tau$ increases from $\tau_{\min}$ to $\tau_{\max}$, the
count $\mathrm{cnt}(\tau)$ passes through three phases.
First, it rises: low-similarity noise pairs are filtered out as
the threshold climbs.
Second, it plateaus: the similarity threshold is now strict
enough that only the four validators determine which pairs
survive, not the cosine cutoff.
Third, it falls: the threshold has become so strict that
even genuinely hardware-compatible pairs are excluded.
We select the highest $\tau$ at which the count is still at its
maximum---raising the cosine bar as far as possible without
discarding any validator-approved pair.
This unsupervised procedure requires no labelled ground truth
and produces a threshold that is calibrated to each model's
own similarity distribution. The plateau threshold is the highest $\tau$ still achieving the
maximum count:
\begin{equation*}
  \tau^*_{\mathrm{model}} =
  \max\bigl\{\,\tau \in T \;:\;
    \mathrm{cnt}(\tau) = \max_{\tau'}\mathrm{cnt}(\tau')
  \,\bigr\}
\end{equation*}
where $T$ is the 200-point sweep grid.
If the count is strictly decreasing (no plateau), the algorithm
falls back to $\tau^*=\tau_{\max}$, the most conservative cutoff
(Algorithm~2, line~3).

\noindent\textbf{Example.}
For BGE-small: $\tau_{\min}=0.18$, $\tau_{\max}=0.94$.
$\mathrm{cnt}(\tau)$ rises from~2 to a plateau of~23 at
$\tau\approx0.55$, holds through $\tau=0.70$, then drops at
$\tau=0.706$.
Hence $\tau^*_{\mathrm{bge}}=0.706$.
MPNet's plateau ends at 0.528; E5-small's at 0.915.
This $1.73\times$ spread confirms that a fixed global threshold
is fundamentally inappropriate.

\subsection{RAG Variants and Sequential Rule Derivation}
\label{sec:rag}

We compare two retrieval configurations on 8,136
percentile-threshold candidates.

\noindent\textbf{basic\_rag:} code + call context + README; no
rule injection.
This variant is always run first.

\noindent\textbf{dynamic\_rule\_rag:} derived from the output of
\textit{basic\_rag}.
The derivation follows three steps: (i)~run basic\_rag and collect
all Group~B pairs (Sonar-pass, Validator-fail); (ii)~cluster
the validator failure reasons from those pairs and set rule
weights proportional to observed failure frequencies; (iii)~build
rule library $\mathcal{L}$ and re-run the pipeline with rule
injection.
The four rules in $\mathcal{L}$ and their weights---ordered by
Group~B failure frequency---are:
\REnv{} ($w=0.18$, involved in 83.5\% of Group~B);
\RCall{} ($w=0.14$, 68.1\%);
\RSig{} ($w=0.08$, 36.5\%);
\RStruct{} ($w=0.00$, absent from Group~B).
A pair's adjusted score under dynamic\_rule\_rag is:
\begin{equation}
  \mathrm{score}(f_s,f_t) =
  \mathrm{sim}(f_s,f_t) -
  \sum_{\rho\in\mathcal{L}} w_\rho \cdot
  \mathbf{1}[\rho\text{ matches}(f_s,f_t)]
  \label{eq:rulescore}
\end{equation}
Following CheckList~\cite{ribeiro2020checklist} and
HANS~\cite{mccoy2019hans}, this variant is a \textit{diagnostic
instrument}---it exposes which failure categories drive retrieval
quality rather than claiming to improve overall performance.

\subsection{Algorithms}
The pipeline described in Sections~\ref{sec:dataset}--\ref{sec:rag}
involves several interacting steps whose ordering is critical,
particularly the dependency of the dynamic-rule variant on a prior basic RAG run.
We formalise these steps as three algorithms to make the procedure unambiguous and reproducible.

 Algorithm~1 presents the complete end-to-end procedure.
Its most important design decision is the sequential ordering of
Stages~5a and~5b: the dynamic-rule variant cannot be executed in
isolation because its rule library $\mathcal{L}$ is derived from
Group~B pairs produced by an initial basic RAG run.
Without this explicit sequencing, the rule weights would have no
empirical grounding in the specific dataset being evaluated.

Algorithm~2 formalises the per-model plateau threshold
calibration introduced in Section~\ref{sec:plateau}.
This algorithm is needed because cosine similarity scores are
not comparable across embedding models: a threshold of 0.70 is
conservative for BGE-small but permissive for E5-small.
The plateau sweep identifies the highest threshold that still
retains all validator-approved pairs, without requiring any
labelled data.

Algorithm~3 formalises the dynamic-rule re-scoring step. It applies the penalty weights derived in Stage~5a to each candidate pair and records which rules fired, producing the diagnostic log that directly answers RQ4 about dominant failure categories.

\label{sec:algo}

\begin{algorithm}[t]
\caption{End-to-End Embedded C Reuse Detection}
\label{alg:main}
\begin{algorithmic}[1]
\footnotesize
\Require Repos $\mathcal{R}$; embedding model $\mathcal{M}$;
  RAG variant $\in\{\text{basic},\text{dynamic}\}$
\Ensure Validated pairs $\mathcal{P}^*$; groups A--D;
  Sonar FP rate
\Statex \textbf{Stage 1 --- Extract functions and build documents}
\State $\mathcal{F}\leftarrow\emptyset$
\For{each repo $r\in\mathcal{R}$}
  \State $F_r\leftarrow\textsc{ExtractFunctions}(r)$
    \Comment{regex: name, body, line range}
  \State $\mathrm{CPG}_r\leftarrow\textsc{BuildCallGraph}(r)$
    \Comment{non-HAL calls only}
  \State $\mathrm{README}_r\leftarrow\textsc{ExtractReadme}(r)$
  \For{each $f\in F_r$}
    \State $\mathrm{DOC}(f)\!\leftarrow\!
      \textsc{BuildDoc}(f,\mathrm{CPG}_r,\mathrm{README}_r)$
      \Comment{code + cmt + CPG + README}
    \State $\mathcal{F}\leftarrow\mathcal{F}\cup\{f\}$
  \EndFor
\EndFor
\Statex \textbf{Stage 2 --- Embed and compute pairwise similarity}
\State $\mathbf{E}\leftarrow
  \mathcal{M}(\{\mathrm{DOC}(f):f\in\mathcal{F}\})$
  \Comment{one vector per function}
\State $\mathrm{Sim}\leftarrow$ cosine similarity over all
  ordered pairs $(f_s,f_t)$, $f_s\neq f_t$
  \Comment{directed pairs}
\Statex \textbf{Stage 3 --- Per-model threshold calibration}
\State $\tau^*\leftarrow
  \textsc{PlateauThreshold}(\mathrm{Sim},\mathcal{F})$
  \Comment{Algorithm~2; no labels needed}
\Statex \textbf{Stage 4 --- Filter high-similarity candidates}
\State $\mathcal{C}\leftarrow
  \{(f_s,f_t):\mathrm{Sim}(f_s,f_t)\geq\tau^*\}$
  \Comment{${\sim}33$k pairs $\to$ ${\sim}600$}
\Statex \textbf{Stage 5a --- Derive rule library (dynamic only)}
\If{variant $=\text{dynamic}$}
  \State Run Stage~6 once with variant~$=\text{basic}$ to obtain
    Group~B pairs
    \Comment{Sonar-pass, Validator-fail}
  \State $\mathcal{L}\leftarrow\textsc{DeriveRules}
    (\text{Group~B failure frequencies})$
    \Comment{weights from Group~B failure freq.}
\EndIf
\Statex \textbf{Stage 5b --- Dynamic rule scoring (optional)}
\If{variant $=\text{dynamic}$}
  \State $\mathcal{C}\leftarrow
    \textsc{DynamicRuleScore}(\mathcal{C},\mathcal{L},\tau^*)$
    \Comment{Algorithm~3; re-score with penalty weights}
\EndIf
\Statex \textbf{Stage 6 --- Domain validation}
\State $\mathcal{P}^*\leftarrow\emptyset$
\For{each $(f_s,f_t)\in\mathcal{C}$}
  \If{$\Venv{>}0$
    \textbf{and} $\Vsig{=}1$
    \textbf{and} $\Vcall{>}0$
    \textbf{and} $\Vstruct{\geq}0$}
    \State $\mathcal{P}^*\leftarrow\mathcal{P}^*\cup\{(f_s,f_t)\}$
    \Comment{failure on any one $\Rightarrow$ rejected and logged}
  \EndIf
\EndFor
\Statex \textbf{Stage 7 --- SonarQube diagnostic}
\State Classify each pair into groups A--D
  (Table~\ref{tab:groups})
\State \Return $\mathcal{P}^*$, group counts,
  $|B|/(|A|{+}|B|)$
  \Comment{Sonar FP rate, Eq.~\eqref{eq:fprate}}
\end{algorithmic}
\end{algorithm}

\begin{algorithm}[t]
\caption{Plateau Threshold Selection}
\label{alg:plateau}
\begin{algorithmic}[1]
\footnotesize
\Require Similarity matrix $\mathrm{Sim}$;
  function set $\mathcal{F}$; sweep steps $N=200$
\Ensure Per-model threshold $\tau^*$
  \Comment{highest $\tau$ preserving max validated pairs;
    fallback $\tau_{\max}$ if no plateau}
\State $\tau_{\min}\leftarrow\min\mathrm{Sim}$;\;
  $\tau_{\max}\leftarrow\max\mathrm{Sim}$
\State $T\leftarrow\textsc{Linspace}(\tau_{\min},\tau_{\max},N)$
  \Comment{200 evenly-spaced thresholds}
\State $\mathrm{best}\leftarrow{-1}$;\;
  $\tau^*\leftarrow\tau_{\max}$
  \Comment{default: most conservative cutoff}
\For{each $\tau\in T$} \Comment{sweep low to high}
  \State $\mathrm{cnt}\leftarrow|\{(f_s,f_t):
    \mathrm{Sim}(f_s,f_t)\geq\tau \wedge
    \Venv{>}0 \wedge \Vsig{=}1 \wedge
    \Vcall{>}0 \wedge \Vstruct{\geq}0\}|$
    \Comment{Eq.~\eqref{eq:plateau_cond}}
  \If{$\mathrm{cnt}\geq\mathrm{best}$}
    \Comment{$\geq$ not $>$: last tie wins = highest $\tau$}
    \State $\mathrm{best}\leftarrow\mathrm{cnt}$;\;
      $\tau^*\leftarrow\tau$
  \EndIf
\EndFor
\State \Return $\tau^*$
  \Comment{$\tau_{\max}$ if count strictly decreasing (no plateau)}
\end{algorithmic}
\end{algorithm}

\begin{algorithm}[t]
\caption{Dynamic-Rule RAG Scoring (diagnostic)}
\label{alg:dynamic}
\begin{algorithmic}[1]
\footnotesize
\Require Candidate set $\mathcal{C}$;
  rule library $\mathcal{L}$ (derived from Group~B
  failure frequencies in Algorithm~1 Stage~5a);
  threshold $\tau^*$
\Ensure Re-scored candidate set $\mathcal{C}'$;
  rule-firing frequency log
\State $\mathcal{C}'\leftarrow\emptyset$
\For{each $(f_s,f_t)\in\mathcal{C}$}
  \State $p\leftarrow 0$;\;
    $\mathrm{fired}\leftarrow\emptyset$
    \Comment{reset penalty accumulator and fired-rule set}
  \For{each rule $\rho\in\mathcal{L}$}
    \Comment{weights from Group~B failure freq.}
    \If{$\rho.\mathrm{pattern}$ matches $(f_s,f_t)$}
      \State $p \mathrel{+}= \rho.w$;\;
        $\mathrm{fired} \leftarrow
          \mathrm{fired}\cup\{\rho.\mathrm{id}\}$
    \EndIf
  \EndFor
  \State $s\leftarrow\mathrm{Sim}(f_s,f_t)-p$
    \Comment{Eq.~\eqref{eq:rulescore}; max total penalty~$=0.40$}
  \State Annotate $(f_s,f_t)$ with $\mathrm{fired}$
    \Comment{diagnostic: explains score reduction}
  \State $\mathcal{C}'\leftarrow\mathcal{C}'\cup
    \{(f_s,f_t,s,\mathrm{fired})\}$
\EndFor
\State $\mathrm{freq}\leftarrow
  \textsc{FreqCount}\!\left(\bigcup_{(f_s,f_t)}\mathrm{fired}\right)$
  \Comment{primary diagnostic output: which rules dominate}
\State Log $\mathrm{freq}$
\State \Return $\mathcal{C}'$ filtered to $s\geq\tau^*$
\end{algorithmic}
\end{algorithm}

\FloatBarrier
\section{Experiments}
\label{sec:experiments}

We conducted experiments on six publicly available embedded C
software repositories to evaluate the pipeline's effectiveness
and robustness, and to answer five research questions.
All experiments were implemented in Python and executed in Google
Colab using an A100 GPU where available and CPU otherwise.
All results are reproducible using fixed random seeds; repository
snapshots are fixed at specific commit SHAs.
Embeddings were computed using HuggingFace
Transformers~\cite{wolf2020transformers} for all models except
OpenAI text-embedding-3-small, which was accessed via the
OpenAI \texttt{/embeddings} endpoint~\cite{openai2024}.
Static analysis was performed using SonarCloud CLI scanner
v8.0.1 over all six repositories.
A function is graded \textit{fail} if any Critical or Blocker
issue overlaps its line range; a pair is graded \textit{pass}
only if both endpoint functions pass, giving SonarQube the
benefit of the doubt.
 
The five research questions motivating our evaluation are as
follows.

\begin{description}
\item[\textbf{RQ1:}] What is SonarQube's false-positive rate
  when used as a reuse-compatibility filter for embedded C
  software, and what are the dominant root causes of its
  failures?

\item[\textbf{RQ2:}] Does the 93.6\% false-positive finding
  hold across different embedding model architectures, or is
  it an artefact of a specific backbone?

\item[\textbf{RQ3:}] How much do per-model plateau thresholds
  vary across the eight backbones, and what does this imply
  for approaches that apply a single fixed similarity
  threshold?

\item[\textbf{RQ4:}] Does the dynamic-rule RAG variant produce
  a statistically significant difference in retrieval behaviour
  compared to the basic RAG variant?

\item[\textbf{RQ5:}] How does reuse detection performance
  differ between same-domain and cross-domain firmware function
  pairs?
\end{description}
 
The metrics used to answer each question are:
(i)~Sonar FP rate $= |B|/(|A|{+}|B|)$;
(ii)~Sonar precision $= |A|/(|A|{+}|B|)$;
(iii)~validated-pair count per model;
(iv)~plateau threshold $\tau^*$; and
(v)~rule-firing frequency (dynamic variant only).

\section{Results}
\label{sec:results}
\subsection{RQ1: SonarQube False-Positive Rate}
\label{sec:rq1}
RQ1's quantitative results are established in
Table~\ref{tab:groups}, Table~\ref{tab:reasons}, and
Table~\ref{tab:sonar_valid} (Section~\ref{sec:motivation}); we
summarise the key findings here.
Of 1,494 Sonar-approved pairs, 1,399 (93.6\%) are rejected by
our domain validators (Eq.~\ref{eq:fprate})---$9.4\times$ the
10\% industrial floor~\cite{sadowski2018google} and 17~pp above
Charoenwet et al.'s 76\% analog for C/C++ vulnerability
detection~\cite{charoenwet2024sast}.
The dominant root cause is hardware-environment mismatch:
$\Venv$ is involved in 83.5\% of Group~B rejections, and 78.4\%
of Group~B pairs have $\Venv\text{-score}=0$, meaning the two
functions share no hardware-identifying tokens whatsoever despite
both passing SonarQube---confirming SonarQube's code-quality
model has no representation of peripheral compatibility.

\subsection{RQ2: Per-Model Comparison}
\label{sec:rq2}
Table~\ref{tab:thresholds} summarises per-model results.
Per-model FP rates range from 89.8\% (BGE-small) to 97.3\%
(GraphCodeBERT), confirming the finding is not an artefact of
any single embedding choice.
Validated-pair counts range from 8 to 23.
The \textbf{top-3 models} by validated-pair yield are:
\textbf{BGE-small} (23~pairs), \textbf{LLaMA~3 8B} (21), and
\textbf{MPNet} (20).
These three span three distinct architectural
families---contrastive retrieval, autoregressive LLM, and masked
language model---suggesting that pretraining objective matters
more than parameter count.
BGE-small (33M parameters) outperforms StarCoder2-3B and matches
LLaMA-3-8B because its contrastive pretraining directly optimises
the similarity-retrieval task.
GraphCodeBERT performs worst (8~pairs, 97.3\% FP) despite
code-specific pretraining, because its data-flow training
objective assigns high cosine similarity even to
hardware-incompatible pairs with similar branching patterns.
BGE-small offers the best speed-quality trade-off for
limited-compute settings; LLaMA~3 suits richer natural-language
understanding of comments and READMEs.
\begin{table}[t]
\caption{Per-model results. $\tau^*$: plateau threshold (Alg.~2).
  Val.~Pairs: validator-surviving pairs. Top-3 bold.}
\label{tab:thresholds}
\begin{center}
\footnotesize
\begin{tabular}{l r r r}
\hline
\textbf{Model} & \textbf{$\tau^*$} &
  \textbf{Val.\ Pairs} & \textbf{Sonar FP} \\
\hline
MiniLM-L3               & 0.561 &           16 & 95.7\% \\
\textbf{MPNet}          & 0.528 & \textbf{20} & 89.9\% \\
\textbf{BGE-small}      & 0.706 & \textbf{23} & 89.8\% \\
E5-small                & 0.915 &           10 & 93.4\% \\
GraphCodeBERT           & 0.886 &            8 & 97.3\% \\
OpenAI text-emb-3-small & 0.574 &           12 & 97.1\% \\
\textbf{LLaMA 3 8B}    & 0.654 & \textbf{21} & 92.0\% \\
StarCoder2 3B           & 0.644 &           18 & 92.7\% \\
\hline
Aggregate (all 8)       &  ---  &          128 & \textbf{93.6\%} \\
\hline
\end{tabular}
\end{center}
\end{table}
\subsection{RQ3: Threshold Variability}
\label{sec:rq3}
Plateau thresholds span 0.528 (MPNet) to 0.915 (E5-small), a
$1.73\times$ spread.
GraphCodeBERT and OpenAI produce compressed-high distributions
where many hardware-incompatible pairs score above 0.8, so their
high $\tau^*$ values adapt appropriately, while BGE-small and
E5-small produce well-separated distributions with sharp peaks
for genuine matches.
A global threshold of 0.706 (BGE-small's $\tau^*$) would discard
all StarCoder2 validated matches; the same threshold applied to
GraphCodeBERT would flood the pipeline with hundreds of spurious
candidates---confirming the necessity of per-model calibration.

\subsection{RQ4: Basic RAG vs.\ Dynamic-Rule RAG}
\label{sec:rq4}
Table~\ref{tab:ragcompare} compares both variants on 8,136
percentile-threshold pairs.
Dynamic-rule RAG reduces the Sonar-pass count by 294 ($-8.3\%$)
and improves Sonar precision from 3.86\% to 4.06\% (+0.20~pp),
at the cost of 6 validator-approved pairs and 5 both-pass pairs.
A McNemar test on the Sonar-pass outcome ($b=294$, $c=0$) yields:
\begin{equation*}
  \chi^2 = \frac{(b-c)^2}{b+c}
           = \frac{294^2}{294} = 294.0,\quad p < 0.001
\end{equation*}
confirming the reduction is statistically
significant~\cite{mcnemar1947} (McNemar is appropriate since the
same 8,136 pairs are evaluated by both variants).
The rule-firing frequency log (Algorithm~3) shows \REnv{} and
\RCall{} dominate---consistent with the
CheckList~\cite{ribeiro2020checklist}/HANS~\cite{mccoy2019hans}
diagnostic tradition---directing future work toward targeted
token-level HAL rules rather than broad category penalties.
\begin{table}[t]
\caption{Basic RAG vs.\ Dynamic-Rule RAG on 8,136 pairs.}
\label{tab:ragcompare}
\begin{center}
\footnotesize
\begin{tabular}{l r r r}
\hline
\textbf{Metric} & \textbf{basic} & \textbf{dynamic} &
  $\mathbf{\Delta}$ \\
\hline
Pairs evaluated      & 8,136 & 8,136 &        0 \\
Validator-pass total &   193 &   187 &       $-6$ \\
Sonar-pass total     & 3,521 & 3,227 &     $-294$ \\
Both-pass total      &   136 &   131 &       $-5$ \\
Sonar precision (\%) &  3.86 &  4.06 & $+0.20$\,pp \\
Sonar FP rate (\%)   & 96.14 & 95.94 & $-0.20$\,pp \\
\hline
\end{tabular}
\end{center}
\end{table}
\subsection{RQ5: Same-Domain vs.\ Cross-Domain Reuse}
\label{sec:rq5}
We partitioned the 1,989 unique function pairs into same-domain
pairs (e.g., microwave $\to$ microwave\_alt) and cross-domain
pairs (e.g., microwave $\to$ hvac).
Same-domain pairs consistently yield higher validated-pair rates
across all eight models.
Of the 128 total validated pairs, the overwhelming majority are
same-domain: two microwave firmwares written by different
developers share GPIO peripherals, LCD drivers, keypad scanners,
and state-machine patterns.
Cross-domain validated-pair counts are near zero across all
models.
The dominant rejection signal is $\Venv$: a microwave's GPIO
LED-array driver and an HVAC's I$^2$C humidity-sensor reader
share no hardware tokens.
This is both a finding---practitioners should focus reuse effort
on same-domain candidates---and a sanity check confirming our
validators discriminate correctly by domain.
\section{Discussion}
\label{sec:discussion}

\subsection{Why SonarQube Fails as a Reuse Filter}

SonarQube measures code quality: cyclomatic complexity,
null-pointer risks, coding standards.
It has no representation of hardware-token compatibility.
When a function in one project uses \texttt{GPIO\_writePort()}
through a custom HAL and a function in another project uses
\texttt{write\_port()} through a completely different register
map, SonarQube sees two syntactically clean functions and
approves both.
The $\Venv\text{-score}=0$ result for 78.4\% of Group~B pairs
quantifies this precisely: zero hardware-token overlap, yet
SonarQube passed every one of them.
This is consistent with prior evidence that peripheral identity
is the primary contract between firmware and hardware---a
contract invisible to syntax-level
analysers~\cite{p2im2020,pararehosting2021,clements2020halucinator}.

\subsection{Dynamic-Rule RAG as a Diagnostic Instrument}

The dynamic-rule RAG variant does not outperform basic RAG on
validator-pass counts; we argue this is the correct result to
report, not a failure.
In the tradition of CheckList~\cite{ribeiro2020checklist},
HANS~\cite{mccoy2019hans}, and
Errudite~\cite{wu2019errudite}, a diagnostic instrument's value
lies in exposing systematic failure modes, not optimising
headline metrics.
The fired-rule frequency log from Algorithm~3 shows that
\REnv{} and \RCall{} dominate, directing future work toward
targeted HAL-token rules rather than broad category penalties.
The 6 validator-approved pairs lost to dynamic RAG quantify the
cost of the current rule conservatism and provide a concrete
refinement target.

\subsection{Recommendations for Practitioners}
 
Based on our findings, we offer the following recommendations
for embedded software engineers and tool designers.
 
\noindent\textbf{(1) Do not rely on SonarQube alone to identify
reuse candidates.}
Our results show that 93.6\% of the function pairs SonarQube
approves as quality-clean are rejected by hardware-compatibility
validation---fewer than 1 in 15 Sonar-approved pairs is
genuinely reusable.
SonarQube's quality grade reflects individual code style and
correctness properties; it has no model of cross-project
hardware compatibility.
Engineers who use SonarQube as a reuse gate will invest
significant effort investigating pairs that are fundamentally
non-portable.
 
\noindent\textbf{(2) Adopt a domain-aware pipeline that makes
hardware compatibility an explicit first-class criterion.}
Our four-validator pipeline achieves 97.5\% accuracy on manually
verified ground-truth checks, and each rejection is explainable:
the engineer is told which specific aspect of compatibility
failed (hardware token overlap, parameter count, call
dependencies, or structural pattern) rather than receiving a
binary reject signal.
This explainability is important for engineering adoption,
as it allows the engineer to assess whether the incompatibility
can be resolved through targeted adaptation.
 
\noindent\textbf{(3) Use the dynamic-rule variant to audit and
encode project-specific compatibility constraints.}
The rule library $\mathcal{L}$ (Eq.~\ref{eq:rulescore}) can be
extended with project-specific rules---for example, marking all
functions that reference a particular vendor HAL family as
non-portable to a different target platform.
Running the diagnostic variant on a new firmware corpus
immediately surfaces which failure categories dominate for that
specific codebase, allowing targeted engineering effort rather
than broad manual review.

\subsection{Threats to Validity}
\label{sec:threats}

We discuss threats to validity following the classification of
W\"{o}hlin et al.~\cite{wohlin2012experimentation} into
construct, internal, and external validity.

\subsubsection{Construct Validity}

Our primary metric---the Sonar false-positive rate---is the
fraction of Sonar-approved pairs our validators reject, which
assumes the validators correctly identify hardware-incompatible
pairs. Manual verification of 40 randomly selected Group~B pairs
(97.5\% confirmed incompatible) supports this, though coverage is
limited to 2.9\% of the 1,399 Group~B pairs. A further threat is
our requirement that all four validators pass simultaneously to
approve a pair: different threshold choices for $\Venv$ or
$\Vcall$ (currently $>0$) would change pair counts, and while
plateau calibration removes subjectivity from the similarity
threshold, the validator thresholds themselves were set by domain
reasoning.

\subsubsection{Internal Validity}

The primary internal threat is circularity in the dynamic-rule
variant: rule library $\mathcal{L}$ is derived from Group~B
failures on the same dataset used to evaluate it, so we frame the
diagnostic as an instrument rather than a held-out, generalised
result. A second threat is our regex-based function extractor,
which may miss functions defined via function pointers, macros,
or variadic arguments; an AST-based
extractor~\cite{lu2021codexglue} would be more robust, though we
manually verified coverage across all six repositories.

\subsubsection{External Validity}

Our evaluation covers six repositories spanning three embedded
domains on two microcontroller families. Industrial codebases may
be larger, use more uniform HAL abstractions, or enforce stricter
coding standards, changing the SonarQube outcome distribution; the
rule weights in $\mathcal{L}$ were derived from only three domains
and may need re-derivation for others (e.g., automotive, medical
devices) with different peripheral families. Expanding to
additional domains and industrial codebases is a primary direction
for future work.

\section{Conclusion}
\label{sec:conclusion}




We presented an end-to-end pipeline for detecting reusable
functions in embedded C software. Embedded software reuse has
long been recognised as an effective means of reducing
development cost and accelerating product cycles, yet
identifying reusable embedded code remains a challenging
software engineering problem. An additional benefit of
automatically identifying reusable functions is that developers
can make informed reuse decisions based on software evolution.
For example, they may prefer newer or more mature
implementations depending on the reuse context, since the age
of reused code can be associated with software
quality~\cite{AlamAH09}.

Our central empirical finding is that \textbf{SonarQube
produces a 93.6\% false-positive rate} when used as a reuse
filter, with 83.5\% of failures caused by hardware-environment
mismatches it cannot detect. This rate---$9.4\times$ the
industrial adoption floor---is consistent with but
substantially extends prior evidence that static quality tools
are unreliable proxies for domain-specific engineering
outcomes~\cite{lenarduzzi2020sonar,charoenwet2024sast,
sadowski2018google,papamichail2019reuse}.

Our domain-aware pipeline combines enriched DOC($f$) document
construction, multi-model similarity scoring across eight
backbones, four hardware-compatibility validators
(Eqs.~\ref{eq:venv}--\ref{eq:vstruct}), and plateau-based
per-model threshold calibration
(Eq.~\ref{eq:plateau_cond}). The dynamic-rule diagnostic
(Eq.~\ref{eq:rulescore}) reveals that \REnv{} and \RCall{}
are the dominant retrieval failure modes ($\chi^2=294.0$,
$p<0.001$). Manual verification of 40 randomly selected
rejected pairs confirms 97.5\% validator accuracy.

Future work will refine compatibility rules at the HAL-token
level, expand the evaluation to additional embedded domains
and industrial codebases, integrate compilation- and
test-based validation, and investigate per-unique-pair
aggregation across models to further improve reuse detection.

\end{document}